\documentclass[american,english]{article}
\usepackage[LGR,T1]{fontenc}
\usepackage[latin9]{inputenc}
\usepackage{geometry}
\usepackage{amstext}
\usepackage{amssymb}
\usepackage{graphicx}
\usepackage{subscript}

\makeatletter

\newcommand{\noun}[1]{\textsc{#1}}
\DeclareRobustCommand{\greektext}{%
  \fontencoding{LGR}\selectfont\def\encodingdefault{LGR}}
\DeclareRobustCommand{\textgreek}[1]{\leavevmode{\greektext #1}}
\ProvideTextCommand{\~}{LGR}[1]{\char126#1}

\makeatother

\usepackage{babel}
\begin{document}
\title{Granular Space--time: The Nature of Time}
\author{Carlton Frederick}
\maketitle
\begin{center}
Central Research Group
\par\end{center}
\begin{abstract}
Granular space-time posits that everything can be expressed as a function
of space-time and matter. And this includes the quantum wave function
\textgreek{Y}. To give a geometric interpretation of \textgreek{Y},
we first need to examine time. The fact that the wave function is
complex results in the time dimension also being complex with the
imaginary component being rolled-up. The symmetry of time is deduced.
\end{abstract}

\section*{Introduction}

A much earlier paper\cite{L1} pointed out that vacuum energy fluctions
were equivalent to mass fluctuations. And those would generate stochastic
fluctuations of the metric tensor. That was interpreted as saying
that the 'points' (events) of space-time moved in relation to other
points. Those mobile points nevertheless continued to tessilate the
space-time manifold

A later paper\cite{L2} noted that a point has no extent and that
would make tessilating of space-time problematic. A granular model
was the proposed. The granules (called 'venues' to distinguish them
from 'events') were taken to be four dimensional; Planck length cubed
by Planck time. The two papers were able to generate many of the results
of quantum mechanics.

The granular model takes as given that all of physics can be described
by properties of space-time and matter. \emph{(and fields and potentials,
but we feel those can also be expressed as properties of space-time
and matter and constraints determined by matter)}. If that is true,
then the solutions of the Schrödinger equation, the wave function,
should also have a geometric representation. To find it, we must first
consider the nature of time.

\section*{Time}

We'll start with a crude idea of time and methodically sculpt it into
a (it is to be hoped) attractive model. We begin with the proposition
that venues are stocastically migrating in all directions; x,y,z and
also t.

\subsection*{'Time Leaves No Tracks'}

Consider the graph (of 1000 points) below. (The vertical and horizontal
lines are artifacts of the graphing software.) The graph represents
the path of a a single venue migrating in x and also in t, where the
coordinate axes are laboratory x and laboratory t.

\includegraphics[scale=0.6]{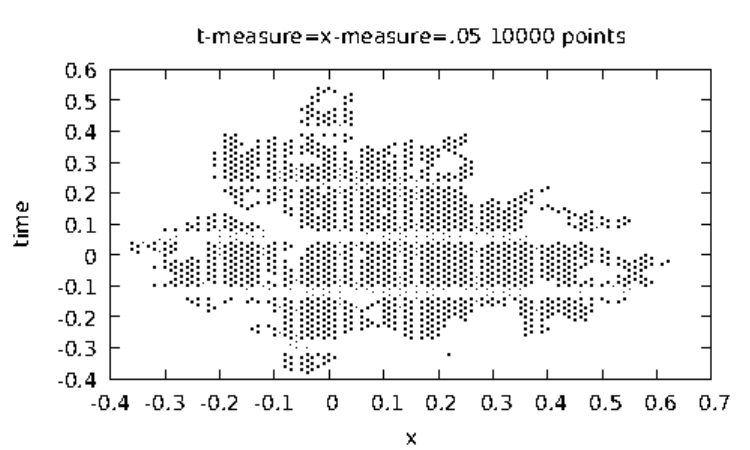}

There is an immediate problem:

Consider what this graph signifies: At any given laboratory-time t,
the same venue will (simultaneously) be at a very large number of
x coordinates. If there were mass/energy at the venue, this would
be very problematic as causality and conservation of mass would be
violated. 

We'd like to treat the time dimension, t, in the same way as we treat
spacial dimensions. But there is a big difference between a space
and time coordinate: Consider the graphic below:

\noindent\begin{minipage}[t]{1\columnwidth}%
\includegraphics[scale=0.68]{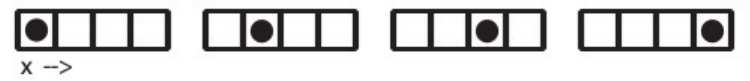}%
\end{minipage}

A particle (the black disk) starts at x=0, then moves to x=1, then
2, then 3. (We are considering space-time to be granular, hence the
coordinate boxes.) There is a single instance of the particle.

But time is different:

\noindent\begin{minipage}[t]{1\columnwidth}%
\includegraphics[scale=0.68]{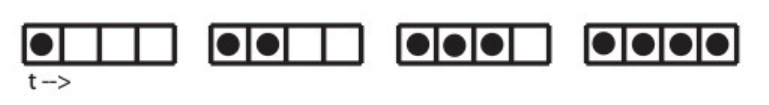}%
\end{minipage}

A particle at rest is at t=0, then moves to t=1, etc. But when it
goes from t=0 to t=1, it also remains at t=0. There are now two instances
of the particle, etc. In other words, a particle at a particular time
is still there as time advances, and the particle is at the advanced
time as well.

We define then, a new quantity, \textgreek{t} (tau-time), that acts
much like the usual time, but in accord with the first graphic, above,
i.e. when the particle advances in time, it erases the previous instance.
That is to say, '$\tau$-Time Leaves No Tracks'. Aside from fixing
the problem of the same mass appearing at an enormous number of different
locations at the same time, \textgreek{t} will be seen to provide
a solution to the collapse of the wave-function problem.

In our model then, the time leaves no tracks concept implies that
there are multiple futures, and they all 'happen'. (This is somewhat
redolent of the Everett many-world interpretation\cite{L6}.) An observation
from the laboratory will select a particular future (making a track).
In general, one can predict the future and also (strange as it sounds)
predict the past.

\subsection*{The Emergence of Time as We Know it}

We assume that venues migrate stochastically in x,y,z,and t. And the
probabilities of migrating in one direction is the same as in the
opposite direction. And this applies to t as well as to x,y,z. So
at this juncture there is no arrow of time. But now consider a very
small symmetry breaking for t, so that the probability of a venue
going forward in time is 50.0000...1 percent and 49.9999...9 percent
probability of going backward in time. This is so close to equal probabilities
as to be effectively equal probabilities.

Now consider two venues bound together by mass. We suggest that it
is mass that causes venues to clump together. In this case, the probabilities
of migrating in one direction are the same as migrating in the other
direction. If this were not the case, the venues (each holding a small
amount of mass) would be ripped apart.

The probability of the venues progressing forward in time versus going
backward is very slightly greater than in the single venue case. This
is easy to see. For example, suppose the single venue probabilities
are two thirds forward and one third backward, a ratio of two to one.
For two venues the probabilities are four ninths forward and one ninth
backward, a ratio of four to one. For each venue added to the clump,
the ratio increases. So that when the clump is large enough to be
non-negligable, e.g. at the quantum particle scale, the arrow of time
(for that clump) is all but completly forward.

Mass then, determines the arrow of time.

There is a problem. When do the migrations occur. Whenness is something
of an amorphous concept when time itself is migrating. A two-component
time (see below) could provide a solution.

\subsection*{The Dual Nature of Time}

'Time' can be considered to have two characteristics: a coordinate
(\ensuremath{\tau}) from minus to plus infinity (or from the big bang
to some end of time), and a sequencer (\ensuremath{\upsilon}), \foreignlanguage{american}{an
ordering schema as described by H. Reichenbach\cite{L10}.} determining
the direction and 'speed' of time.

\selectlanguage{american}%
So for migrations in time as well as space, we need two kinds of time:
coordinate time (as in x,y,z, and t), and sequential time (a measure
of something coming before or after something else, and the interval
between them).

Our approach will be to decompose time (total time, \includegraphics[scale=0.2]{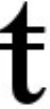}
which we call t-bar) into coordinate time $t_{c}$ and sequential
time $\tau$. We'll model sequential time, and then combine the two
times back into total time.

We consider coordinate time in the usual way, a linear coordinate
that runs from minus to plus infinity. But, since there are posited
to be no times shorter than a Planck time, instead of a mathematical
line, coordinate time is an ordered set of points one Planck time
apart. The sequencer implies that all venues are in constant motion.
\emph{Omnia mutantur} (all things change) as Ovid puts it.

For sequencing, we merely need to know what event comes before or
after another (i.e. the sequence and whether the sequence goes up
or down), and the time interval (a single Planck time). A section
of a line would do. But there doesn't seem to be any use for a zero
of the sequence, nor infinity either. The zero seems arbitrary. (With
zero and infinity, the sequence would be more like a coordinate.)
So this suggests a circle rather than a line-{}-an angle measure would
suffice, i.e. a phase.

\includegraphics[scale=0.6]{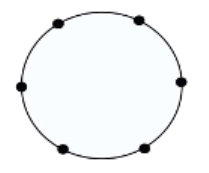}$\tau$ time

There is a problem with the above. If the radius is $t_{p}$ the 'arcs'
between the sequence points, also $t_{p}$, will not overlap as the
sequence advances more than one circumference. (i.e. in the above
diagram, if the radius is one Planck time, the 'arcs' between the
points can't be one Planck time.) And we'll need that overlap. At
this scale though, we cannot really define a circle as there can be
no 'arcs' at the Planck scale. But we can define the equivalent of
a circle, as follows:

\includegraphics[scale=0.6]{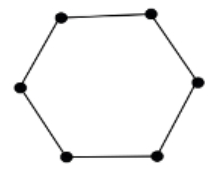}$\tau$ time (a better representation). 

And with this regular hexagon, the 'arc' lengths equal the radius,
so we get overlap. (And here $\pi_{p}$, Planck pi, is a convenient
3.0)

We will presently represent time as \includegraphics[scale=0.2]{tbarsmall}$=t_{c}e^{i\tau}$
where $\tau$ is sequence time. (We could then represent \includegraphics[scale=0.2]{tbarsmall}
squared, as for instance in $ds^{2}$ as \includegraphics[scale=0.2]{tbarsmall}\textsuperscript{{*}}\includegraphics[scale=0.2]{tbarsmall}.)\footnote{\noun{note:} Imaginary time is not unheard of in quantum mechanics:
The Diffusion equation with imaginary time becomes the Schrödinger
equation. And our QM model is diffusion of space-time (venues). For
Stephen Hawking no real distinction exists between \textquoteleft real\textquoteright{}
and \textquoteleft imaginary\textquoteright{} time. He recommends,
in line with his instrumentalist philosophy, that we should adopt
a notion of time, which leads to the best description of physical
reality (quoted in ``The March of Time'' Friedel Weinert, Springer
2013). Imaginary time has also been explored by Neil Turok (Quantum
Theory: A Two-Time Success Story, Springer, 2014) and and Dmitri Sokolovski
(Time in Quantum Mechanics, Vol 1, Springer 2008)}

But while we'll be using complex time, a purely real representation
has some conceptual advantages. Consider the following: (The idea
is to recombine sequence and coordinate times into one real coordinate.)

\includegraphics[scale=0.5]{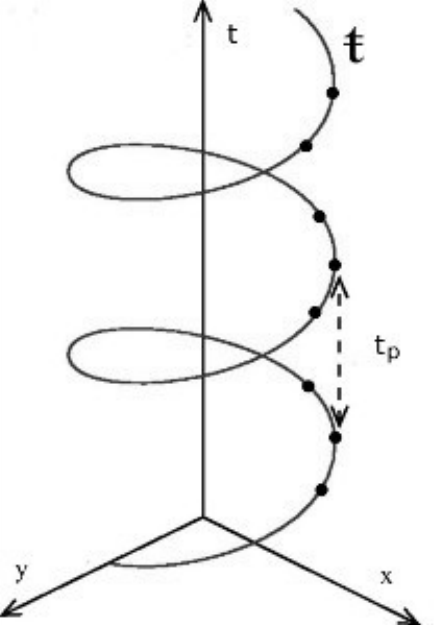}

We could represent \includegraphics[scale=0.2]{tbarsmall}=(t\textsubscript{c},\textgreek{t})
as a single time coordinate, not as a line, but as a helix. $t_{c}$
is the center of the helix, here marked as t (the coordinate for time
in the macro-world). The 'rungs' of the helix are one Planck time
apart. Whenever there is a time migration, time goes either up or
down one rung, leaving the sequencing unchanged. The helix should
be considered, not as circular in nature, but hexagonal as in the
above representation of $\tau$ (sequence) time. Further, the helix
is 4-dimensional as the above diagram is the same in the x,z and y,z
planes. (Re. imaginary time: note that the above helix can be generated
using $e^{ix}$ with x real. The value of x and the real and imaginary
parts of $e^{ix}$ provide the three real dimensions.) These helixes
are suggestive as we will later consider an aggregate of venues (holding
a mass) as being a helical distortion of space-time i.e. the wave
function wave.

With two types of times, we also have two types of time reversal.
Coordinate time reversal is the conventional notion: running the movie
backwards. Reversal of sequence time, we argue, as it governs sequencing,
results in the complementing of the space measures .

\selectlanguage{english}%
We now return to our preferred model of time as a complex dimension.
(We do this because the Schrodinger wave equation's solutions are
complex and we consider the waves as purely a distortion of space-time-matter.
The wave equation must be complex (among other reasons) to assure
that the free particle (complex) wave can have a constant probability
density (\textgreek{Y}\textsuperscript{{*}}\textgreek{Y}).

We'll consider the sequencer function to be described by the imaginary
(rolled-up) component of time (the phase). We note that a rolled up
(real) fifth dimension was postulated by Kaluza and Klein\cite{L13}
to bring electromagnetism into the metric tensor. We have shown that
we can optain the Kaluza result using a rolled up imaginary fourth
dimension.

The imaginary component acts much like a separate (time-like) fifth
dimension. This is vaguely similar to the idea that there is a fifth
dimension which \emph{is} mass, as proposed by Mashhoon \& Wesson\cite{L11}
and the Space-Time-Matter consortium \cite{L12}.

We define Total (complex) time T.

T=\ensuremath{\tau}+i\ensuremath{\upsilon}.

\ensuremath{\tau} is the 'Time Leaves no Tracks' version of t.

\ensuremath{\upsilon} is the imaginary component of time. It is rolled-up
at the Planck scale so in the macroworld T is indistinguishable from
\ensuremath{\tau}.

(Note: 'complex time' is not an entirely new idea, e.g. S. Hawking\cite{L14}.)

A property of time is that it (usually) advances. As \ensuremath{\upsilon}
is a component of time, we assume it advances as well. But \ensuremath{\upsilon}
is rolled-up, so, as it continuously advances, it continuously reaches
a maximum and rolls over to zero. We represent this as a frequency.

Masreliez\cite{L15} and Mukhopadhyay\cite{L16} among others have
suggested that a mass oscillates at its Compton frequency, (and without
such oscillation, there would be no DeBroglie wave, or indeed a \textgreek{Y}).
We accept that suggestion. The Compton frequency ,$f_{c}$, is defined
as $f_{c}=\frac{mc^{2}}{h}Hz$. 

We first convert Hz to cycles/Planck time.

$\frac{f_{c}}{\sqrt{\frac{hG}{c^{5}}}}=\frac{mc^{2}}{h}$ 

Now we'll convert m from kilograms to Planck mass, $m_{p}$.

$\frac{f_{c}}{\sqrt{\frac{hG}{c^{5}}}}=\frac{mc^{2}}{h}\sqrt{\frac{hc}{G}}$

Simplifying, we have $f_{c}=m_{p}$. 

This says that if the mass in a venue is zero, (from the viewpoint
of the laboratory observer) the \ensuremath{\upsilon} time does not
advance. The more mass in a venue, the more 'rapidly' \ensuremath{\upsilon}
advances until at a maximum venue mass of one Planck mass, the frequency
has increased to one cycle per Planck time. And in that latter case,
every Planck time advances \ensuremath{\upsilon} to the same angular
point, which is then indistinguishable from a frequency of zero. In
short then, we associate mass with a frequency (the Compton frequency)
of the imaginary time component.

\selectlanguage{american}%
We consider now, a space-time occupied by a single (indivisible, a
quark perhaps) mass. 

As a trial idea, suppose a venue with mass migrates in all directions
including time, i.e. it also migrates in (sequence) time.

Time on average moves forward for one Planck time, and then on average
backward for one Planck time, etc. The average wavelength then, is
2$t_{p}$seconds and the frequency f is 1/(2$t_{p}$) Hz.

The Compton frequency ($f_{c}$) is defined as $f_{c}=\frac{mc^{2}}{h}$.
Equating the two frequencies gives,

$\text{1}=\frac{1}{m}(\frac{h}{2c^{2}t_{p}})$.

But the above is for only one space dimension. If we consider three
space dimensions, then whenever an angle migration is due to occur,
three migrations must happen (one for each angle axis). They can't
happen at the same time as the rotation group is non-commutative.So
in the above equation we must replace the above with,

3 $=\frac{1}{m}(\frac{h}{6c^{2}t_{p}})$.

Substituting for h, c, and $t_{p}$ yields $m\cong2.43*10^{-8}kg$.

This is a remarkable result as the mass is very close to the Planck
mass.

Instead of 1 or 3 above, we now ask the value for exactly one Planck
mass, and substitute,

$t_{p}\equiv\sqrt{\frac{hG}{2\pi c^{5}}}$, and $m_{p}\equiv\sqrt{\frac{hc}{2\pi G}}$
where G is the constant of gravitation, we obtain,

$\pi/3$ $\thickapprox$1.047, a value slightly greater than unity.
Perhaps then, we should use the Planck pi, $\pi_{p}$, rather than
$\pi$. And since $\pi_{p}$$=3$, that gives a value of exactly one.

This is a nice result as it suggests:

\quad{}1-the Planck length is the smallest possible length,

\quad{}2-the Planck time is the smallest possible time, and

\quad{}3-the Planck mass is the smallest possible classical mass
(i.e. not subject to quantum mechanics).

In contrast to conventional QM where a massive particle has a wave
function this model predicts that there isn't a wave function for
a sufficiently large, mass. So, even in principle, the two slit experiment
cannot be done with marbles, or cannonballs.

As to \foreignlanguage{english}{\textgreek{Y}}\textsuperscript{\selectlanguage{english}%
{*}\selectlanguage{american}%
}\foreignlanguage{english}{\textgreek{Y} We can still consider a probability
curve but we'll interpret it differently: If we take any (horizontal)
time (\textgreek{t}) as a 'now', A venue (containing a mass) stochastically
flits forward and back in space-time. So that at 'now' there is one
and only one particle. But where it is cannot be predicted. However,
the likelihood of the particle being at a particular x (+/- dx) position
is determined by the relative number of times the particle is at that
position. This is analogous to \textgreek{Y}}\textsuperscript{\selectlanguage{english}%
{*}\selectlanguage{american}%
}\foreignlanguage{english}{\textgreek{Y}. But the probability curve
is a construct. It represents, but is not actually, the particle.
When the particle is measured by, for example, being absorbed in a
detector, it freezes (no longer moves stochastically). It no longer
flits through time and space so the graph 'collapses' to the measured
position. (that position is only determinable by the measurement.)
This is analogous to the collapse of the wave function, but here there
is no collapse problem. A previous paper \cite{LX1}noted that probability
is observer dependent, and that applies to \textgreek{Y}}\textsuperscript{\selectlanguage{english}%
{*}\selectlanguage{american}%
}\foreignlanguage{english}{\textgreek{Y} as well.}

\selectlanguage{english}%
There are a few points/speculations to be made about measurements.
First, to be a true measurement, there must be a latch/flip-flop/memory
so that the 'film' cannot be run backwards. As an example, consider
the two slit experiment with electrons. If a measurement device is
placed at a slit, there is no interference pattern. But when an electron
goes through a slit, the orbital electrons in atoms of the wall of
the slit will be distorted by the passage of the electron. This distortion
is \emph{almost} a measurement. But when the electron passes through
the slit, the orbital electrons become un-distorted. The interference
pattern is still produced because there is no latching of measurement
information. A latch could be some mechanical contrivance, or even
human (or non-human) memory. A fruit-fly observing at the slit will
kill the interference pattern, but only for the fruit-fly. We think
the process should be transitive; A human observing the fruit-fly's
memory will cause the interference to be killed for the human as well.
A measurement forges a connection between the thing being measured
and the measurer--forcing them to have the same relative now. In
the macro-world, virtually everything observes (via photons) everything
else, forcing that macro-world (or a portion thereof) to have the
same relative now. And measurements forces time to have tracks. Not
that time is frozen, but looking back to a particular time will show
uniquely what the world looked like at that time. E.g., if one were
to do high-speed filming of particle 'tracks' in a cloud chamber,
one would see the time-tracks.

Observation, a crucial part of a measurement, is conducted via photons.
We speculate that \emph{all} measurements are via photons.

The time leaves no tracks concept implies that there are multiple
futures, and they all 'happen'. (This is somewhat redolent of the
Everett many-world interpretation\cite{L3}) In Stochastic Granular
Spave-time theory, an observation from the laboratory will select
a particular future (making a track).

Let's consider the idea of the 'world-line'. Moving forward from the
present, we are predicting the future. And with quantum uncertainties
(as well as with the intervention of outside forces) that future cannot
be predicted according to classical determinism. And if there is no
completely deterministic trajectory going forward, then arguably neither
is there one going backward in time. The world-line then, seems to
have limited utility in quantum mechanics. Instead of a world-line,
we consider a 'world-double-cone', with its apex at 'now' that widens
as one moves forward or backward in time. So while quantum mechanics
lets us probabilistically predict the future it also lets us probabilistically
predict the past. 

\section*{(Brief) Discussion}

Stochastic Granular Space-time is neither String Theory nor Loop Quantum
Gravity but is concerned with aspects common to both. In particular
the nature of time is a thread running through the three theories.
This paper is concerned mainly with time.

In order that we treat time in the same way as we treat space (and
not to have particles appear at different places at the same time),
we needed a new interpretation of time: 'Time Leaves No Tracks', complex
time with the imaginary component 'rolled-up' at the Planck scale,
and a very slow breaking of the symmetry of time (the arrow of time),
and possibly a connection between mass and time (suggested by the
Schwartschild solution). The implication is that our usual notion
of time is just a human construct, not actually intrinsic to space-time. 

\section*{Acknowlegements}

I thank Nick Taylor for most helpful discussions. And especially,
I thank Lee Smolin and Carlo Rovelli for their illuminating books\cite{L8,L9}
on Loop Quantum Gravity.

\end{document}